\documentstyle[preprint,prc,aps]{revtex}
0
\begin{document}
\draft
\def\relR{\mbox{\small \bf R}}
\def\inr{\mbox{\small \bf r}}
\def\opera{\mbox{\large a}}
\def\larg{\mbox\large g}
\def\Dt{\Delta t}
\def\bR{\bar{\relR}}
\def\bt{\bar t}
\def\tn{\tilde n}

\title{Friction coefficient for  deep-inelastic heavy-ion collisions.}
\author{G.G.Adamian$^{1,2}$, R.V.Jolos$^{1}$  and A.K. Nasirov$^{1,2}$}
\address{${}^1$Joint Institute for 
Nuclear Research, Dubna, 141980 Russia}
\author{A.I.Muminov}  
\address{${}^2$Heavy Ion Physics Department, Institute of Nuclear Physics
702132 Ulugbek, Tashkent, Uzbekistan}  
\date{\today}
\maketitle

\begin{abstract}

Based on the microscopic model, the friction coefficient for the  relative 
motion of nuclei in deep-inelastic  heavy-ion collisions is calculated. 
The radial dependence of the  friction coefficient is studied  and
the results are compared with those found by other methods. 
Based on this result, it was  demonstrated that the kinetic energy 
dissipation in deep-inelastic  heavy-ion collisions is a gradual process 
which takes up a significant part of a reaction time.
An advantage of the suggested method is that it allows one to consider   
the relative motion of nuclei and the intrinsic motion self-consistently.
\end{abstract}


\section{Introduction}

Nuclear friction is an important ingredient of theoretical approaches to a 
variety of  nuclear physics phenomena, such as dynamic thresholds for 
compound nucleus formation \cite{1,2,3,4,5},  
enhancement of neutron emission prior to fission \cite{6,7}, 
width of mass and charge distributions in deep-inelastic heavy-ion reactions
\cite{8},  and the width of giant  resonances \cite{9}.
There are  many  experimental results on  deep-inelastic heavy-ion 
collisions (DIC) and fusion-fission reactions which need the introduction
of the nuclear friction concept for their interpretation.
This stresses the importance  of  understanding the nature of 
nuclear friction.

The present paper is devoted to  calculations of the friction 
coefficient for DIC heavy-ion collisions. 
Its appearance is stimulated not only by the 
possibility to perform more exact calculations than earlier,
but also by the new experimental results which
require a more detailed microscopic theory
for their interpretation than  was necessary before.

Different theoretical approaches to this problem are known. 
The majority  of them 
are based on the assumption that the dissipative mechanism is of a one-body
nature \cite{10,11,12}. These models  differ in the structure 
of the intrinsic excitations that are taken into account. These can be pure 
collective excited states \cite{13,14,15,16} or 
incoherent particle-hole excitations \cite{10,18,19}.
Some of the models do not include  nucleon exchange and consider only 
particle-hole excitations, with both the  particle and the hole belonging 
to the same nucleus \cite{20} or vice versa \cite{10}.  
The models also differ in the approximations they use for  including
the finite decay time of  one particle-one hole (1p-1h) excitations in more 
complicated configurations (2p-2h, and so on). Many approaches 
\cite{12,13,18,21}  implicitly use the statistical assumption 
of  rapid  equilibration of the noncollective intrinsic degrees of freedom 
and,  therefore are not applicable to the description of the  initial  phase 
of the reactions, where the main part of kinetic energy dissipation 
takes place. 

The contribution of the actual 1p-1h state or more complicated ones to the 
dissipation process depends on the occupation numbers of the single-particle 
states and their evolution during the reaction. However, in the 
calculations of the nuclear friction coefficient performed up to now, 
the statistical  assumption on  the excitation energy distribution was
realized in the usual way,  meaning the  introduction of temperature and,
correspondingly, of the Fermi occupation numbers  at the very
beginning of the reaction. In principle, temperature introduced
in this way is a time-dependent quantity. However, in practice 
temperature was considered to be a constant  corresponding to the total 
excitation energy \cite{15,17}. Thus, the  time  dependence of the 
single-particle occupation numbers was not taken into account.  Only in the 
approach based on the Dissipative Diabatic Dynamics (DDD) \cite{22,23}
was the evolution of the single-particle occupation numbers 
taken into  account, but under the assumption of diabaticity. 
There are some doubts  however, about the validity of the DDD concept. 
It is  also  known  from the calculations of inelastic processes in 
nucleus-nucleus collisions that appreciable energy dissipation takes place 
even before the first crossing of the single-particle levels near the Fermi 
surface \cite{24}.

We should  also mention  the approach suggested in \cite{25}, where 
relative and  intrinsic motion were consistently treated in a 
time-dependent theory of heavy-ion collisions. The authors presumed 
neither weak coupling between the relative motion and the 
intrinsic excitations nor the canonical distribution function for the
density operator of intrinsic motion. However, the analytical 
expressions for the friction tensor and other characteristics
of the energy transport obtained in \cite{25} were not
applied to calculate them.
         
Thus, it is the aim of the present paper to take into account the  time
evolution of the single-particle occupation numbers during the reaction
by numerical solution of the master equation for them and, based on this 
result, to perform  calculations for the friction 
coefficient. Since the occupation numbers found in this way
correspond to the current kinetic energy losses, this means that
the relative and intrinsic motions are considered self-consistently.
Our model makes it possible to take into account explicitly the influence 
of the nuclear shell structure on the  collision process. Moreover,
we improved the single-particle approximation by a phenomenological 
allowance for the  residual interaction which is treated in the so called
$\tau$-approximation. The radial friction coefficient is calculated 
as a function of the mass and charge of the reaction participants.

The general formalism is given in Section 2. The results of the 
calculations are  presented in Section 3. A summary is given in  
Section 4.

\section{Basic formalism}

It is convenient to start with the total Hamiltonian of a dinuclear system
written in the form 
\begin{equation}
\label{iniham}
\hat  H=\hat H_{rel}(\relR; {\bf P})+\hat H_{in}(\xi)+\delta \hat V(\relR,\xi),
\end{equation}
where  the Hamiltonian of a relative motion,
\begin{equation}
\label{Hrel}
\hat H_{rel}(\relR; {\bf P})=\frac{{\bf \hat P}^2}{2\mu }+
\hat {\cal V}(\hat{\relR}),
\end{equation}
consists of the kinetic energy operator and the nucleus-nucleus interaction
potential $\hat  {\cal V}(\hat {\relR})$. 
Here, $\hat {\relR}$ is the relative distance between
the centers of mass of the fragments, ${\bf \hat P}$ is the conjugate
momentum, and $\mu $ is the reduced mass of the system; $\xi$ is a set
 of relevant intrinsic variables. The last two terms in
(\ref{iniham}) describe the internal motion of nuclei and the coupling between
the relative and internal motions (for details, see \cite{26,27}).
It is clear that the coupling term leads to a dissipation of the kinetic 
energy  into the energy of internal nucleon motion. 

Neglecting at the moment the residual nucleon-nucleon interaction,
whose effect will be included later, we take a sum of the 
last two terms (\ref{iniham}) as a single-particle Hamiltonian of a 
dinuclear system
\begin{eqnarray}
\label{DNSHam}\hat H_{in}(\xi)+\delta \hat V(\relR,\xi)&=&
\hat {\cal H}(\relR(t),\xi)+h_{residual}, \nonumber\\
              \hat {\cal H}(\relR(t))&=&
\sum_{i=1}^{A}\left( \frac{-\hbar ^2}{2m}\Delta _i+
 \hat  V_P(\inr_i-\relR(t)) + \hat V_T(\inr_i)\right), 
\end{eqnarray}
where $m$ is the nucleon mass and $A=A_P+A_T$ is the total number of
nucleons in the system. 

Then,  in the second quantization representation, the Hamiltonian
$\hat {\cal H}(\relR(t),\xi)$ can be written as 
\begin{equation}
\label{DNSsec}
\hat {\cal H}(\relR(t),\xi)=
\sum_{P} \varepsilon_P 
 \opera_P^+\opera_P^{} +
\sum_{T} \varepsilon_T 
 \opera_T^+\opera_T^{}+
 \sum_{i\ne i'} V_{ii'}(\relR(t)) \opera_i^+\opera_{i'}^{}, 
\end{equation}
where
\begin{eqnarray}
\label{coupt} && \sum_{i\ne i'} V_{ii'}(\relR(t)) \opera_i^+\opera_{i'}^{}
=\sum_ {P\ne P'} \Lambda^{(T)}_{PP'}(\relR(t))\opera_P^+\opera_{P'}^{} +  
\sum_{T\ne T'} \Lambda^{(P)}_{TT'}(\relR (t))\opera_T^+\opera_{T'}^{} + \\
&&\sum_{T,P} \mbox{\large g}_{PT}(\relR (t))(\opera_P^+
\opera_T^{} + {\rm h.c.})\,.
\nonumber
\end{eqnarray}
Here  $P\equiv (n_P,j_P,l_P,m_P)$ and $T\equiv (n_T,j_T,l_T,m_T)$ are 
the sets of quantum numbers characterizing the single-particle state in an
isolated  projectile and the target nuclei, respectively. The
single-particle basis is constructed from the asymptotic wave vectors of 
the single-particle states of the noninteracting nuclei-the projectile ion 
$|P>$ and the target nucleus $|T>$ in the form 
\begin{mathletters}
\label{eq:all} %note location
\begin{eqnarray}
|\tilde P>&=&|P>-\frac 12 \sum_T |T><T|P>,\label{eq:a} \\
|\tilde T>&=&|T>-\frac 12 \sum_P |P><P|T>. \label{eq:b}
\end{eqnarray}
\end{mathletters}
For this basis set, the orthogonality condition is satisfied up to terms
linear in $<P|T>$.
Then 
\begin{mathletters}
\label{matrel:all} %note location
\begin{eqnarray}
\Lambda^{(T)}_{PP'}(\relR(t))  =  <P|V_T(\inr)|P'>, \label{matrel:a}\\
\Lambda^{(P)}_{TT'}(\relR(t))  =  <T|V_P(\inr-\relR(t))|T'>, 
\label{matrel:b} \\
\larg_{PT}(\relR(t)) = {1\over2} <P|V_P (\inr-
\relR(t)) +V_T (\inr)|T>. \label{matrel:c}
\end{eqnarray}
\end{mathletters}
The  nondiagonal matrix elements $\Lambda _{PP^{\prime }}^{(T)}
(\Lambda _{TT^{\prime }}^{(P)})$ generate the particle-hole transitions 
in the projectile (target) nucleus. The matrix elements $\larg_{PT}$
are responsible for the nucleon exchange between reaction partners. 
These matrix elements were calculated using the approach proposed in 
\cite{28,29}.  In (\ref{DNSsec}), $\varepsilon_{P(T)}$ are the 
single-particle energies of nonperturbed states in the projectile (target) 
nucleus. The coupling between the intrinsic nuclear degrees of freedom and 
the collective variable $\relR$ is introduced by the $\relR$ dependence of 
the sum of the single-particle potentials in (\ref{DNSHam}). 
Since the trajectory calculation  shows that the relative distance
$\relR(t)$ between the  centers of the interacting nuclei could not be 
less than the sum of their radii the  tail of the partner single-particle 
potentials can be considered as a perturbation disturbing the asymptotic 
single-particle wave functions and their energies.

It is convenient to include the diagonal matrix elements of 
$V_{ii'}(\relR(t))$ in $H_{in}$, introducing the renormalized 
$\relR(t)$-dependent  single-particle energies

\begin{mathletters}
\label{ener:all} %note location
\begin{eqnarray}
\tilde {\varepsilon}_P (\relR(t))  =  \varepsilon_P +
<P|V_T(\inr)|P>, \label{ener:a}\\
\tilde {\varepsilon}_T (\relR(t))  =  \varepsilon_T + 
<T|V_P(\inr-\relR(t))|T>.\label{ener:b}
\end{eqnarray}
\end{mathletters}

When the nuclear forces begin to act between the colliding nuclei, the
velocity of their relative motion can be considered as a small quantity
compared to the Fermi velocity.
Then the speed of the nucleons is  mainly associated with their 
intrinsic motion. Since the relative (collective) motion is rather
slow compared to the intrinsic motion, the perturbation of the 
intrinsic motion produced by  changing the  coupling to the relative 
motion ($\bR$)  can be assumed to be small during some small time 
interval $\Dt$ of an arbitrarily chosen time $t$ \cite{13}. The small 
parameter in our consideration $\Dt$ thus characterizes the time interval 
during which the $\bR$-dependent mean field of the combined dinuclear system 
changes so little that we can neglect the effect of this changing on the 
intrinsic motion. At the same time,  the characteristic time
$\Dt$ can not be taken smaller than the relaxation time of the mean field. 
The situation described above is  suitable for  applying  the 
linear response theory to a description of dissipative heavy-ion 
collisions  \cite{13}.
For this reason, we start from the expression for the friction 
coefficient of the radial motion obtained in that approach  \cite{13},
\begin{equation}
\label{frcoef}
\gamma_{RR}(R(t))=\sum_{ik}\left|\frac{\partial V_{ik}
(\mbox{\small\bf R})}{\partial R}\right|^2
B_{ik}^{(1)}(t).
\end{equation}
\begin{eqnarray}
B_{ik}^{(n)}(t)
&=&\frac{2}{\hbar}\int_{t-\Delta t}^{t}dt'\frac{(t'-t)^n}{n!} \exp
\left(\frac{t'-t}{\tau_{ik}}\right)\sin [{\tilde \omega}
_{ki}({\bf R}(t'))(t-t')]\nonumber \\
&\times& [\tilde n_k(t')-\tilde n_i(t')],
\end{eqnarray}

\noindent
where $ \tau_{ij}=\tau_i \tau_k/(\tau_i+\tau_k)$;
$\tau_i$ is the parameter describing  the damping of the single-particle 
motion. The expression for $\tau_i$ is derived in the theory of quantum 
liquids \cite{30,31} (see Appendix A); 
$\hbar\omega_{ij}=\tilde {\varepsilon}_i (\bR) - 
\tilde {\varepsilon}_j (\bR) $ is the energy of the single-particle 
transition in one of the nuclei as well as between the interacting nuclei.
The important ingredients of this formula are the occupation numbers
of the single-particle states $\tilde n_i(t)$. Since the excitation energy 
of the interacting nuclei changes significantly during the course of the 
collision, it is necessary to take into account the time dependence of 
the occupation numbers. The importance of this point was already stressed 
in \cite{13}. At the same time,  new experimental
data indicate that the assumption of the fast statistical equilibration of
the excitation energy during the collision time, {\it i.e.}, an introduction 
of a time-dependent temperature and Fermi occupation numbers is not adequate
for the physical picture. As  already mentioned in the 
Introduction,  the calculations  of $\gamma_{RR}$ performed up to now 
have been done under the assumption  that the occupation numbers  can 
be taken as the Fermi occupation numbers
$$n_j=(1+\exp((E_j-\lambda)/\Theta))^{-1},$$
where $\Theta$ is the temperature corresponding to the total excitation 
energy of a dinuclear system.

To find the time-dependent occupation numbers $n_j(t)$, we developed
in \cite{26,27,32} a method which is described
briefly  below for completeness of the presentation. 

Since explicit allowance for the residual interaction requires extensive
calculations, it is customary to take   the two-particle collision
integral into account in linearized form ($\tau$-approximation).

Then, the equation for the single-particle density matrix $\tn$ takes 
the form
\begin{eqnarray}
\label{eqoccnum}
i\hbar \frac {\partial \hat{\tn}(t)}{\partial t} =
[\hat {\cal H}(\mbox{\small\bf R}(t)),\hat {\tn}(t)] -
 \frac {i\hbar}{\tau}[\hat {\tn}(t) -
 \hat {\tn}^{eq}(\mbox{\small\bf R}(t))]\,,
\end{eqnarray}
where 
$\tn^{eq}(\relR(t))$ is a local quasi-equilibrium 
distribution, {\it i.e.} a Fermi distribution with the temperature $T(t)$ 
corresponding to the  excitation energy at the internuclear distance
$\bR(t)$. Substituting our Hamiltonian (\ref{DNSsec}) into (\ref{eqoccnum},
we have
\begin{eqnarray}
\label{diagn}
i\hbar \frac {\partial \hat{\tn}_i(t)}{\partial t} &=&
\sum_k \left[V_{ik}(\bR(t))\tn_{ki}(t)-
V_{ki}(\bR(t))\tn_{ik}(t)\right]\nonumber\\
&-&\frac{i\hbar}{\tau_i}\left[{\tn}_i(t)-{\tn}_i^{eq}(t)\right]\,,
\end{eqnarray}
where ${\tn}_i$ is a diagonal matrix element of the density matrix.
The approximate equation for nondiagonal matrix elements takes the form
\begin{eqnarray}
\label{nondiag}
i\hbar \frac {\partial \hat {\tn}_{ik}(t)}{\partial t} &=&
\hbar \left[{\tilde \omega}_{ik}(\bR(t))-\frac{2i}{\tau_{ik}}\right]
\tn_{ik}(t)\nonumber\\
&+&V_{ki}(\bR(t))\left[\tn_k(t)-\tn_i(t)\right],
\end{eqnarray}
where we have used the notations 
${\tilde \omega}_{ik}=\left[{\tilde \varepsilon}_i-
{\tilde \varepsilon}_k\right]/\hbar.$

Assuming incoherence in the phases of the nondiagonal matrix
elements, we use  the following approximation to simplify   
equation (\ref{nondiag}):
$$\sum_{k'}V_{k'i}(\bR(t))n_{k'k}(t)
-\sum_{i'}V_{ki'}n_{ii'}(t)\approx V_{ki}(\bR(t))
\left[n_k(t)-n_i(t)\right].$$
As  formulated above, we shall consider the solution of equations
(\ref{diagn}) and (\ref{nondiag}) for  a small time interval $\Dt$ of
an arbitrarily choosen time $t$. Then the solution of  equation 
(\ref{nondiag}) can be written as
\begin{eqnarray}
\tilde n_{ik}(\bar t)&=& {1\over
i\hbar}\int\limits_{t}^{\bar t}dt'V_{ik}(\mbox{\small\bf R}(t')) \exp
\left\{i~\int\limits_{t'}^{\bar t}dt''\left[\tilde \omega_{ki}(\mbox{\small\bf
R}(t'') + {i\over \tau_{ik}}\right]\right\}\nonumber\\
&\times& [\tilde n_k(t')-\tilde n_i(t')]\,\,,
%\tn_{ik}(\bt)&=&\frac 1{i\hbar}\int_t^{\bt}dt'V_{ik}(\bR(t'))
%\exp\left\{i\int_{t'}^{\bt}dt''\left[\tilde\omega_{ki}(\bR(t''))
%+\frac{2i}{\tau_{ki}}\right]\right\}\nonumber\\
%\times(\tn_k(t')-\tn_i(t')),
\end{eqnarray}
where $t\leq\bt\leq t+\Dt$. Substituting this result into 
Eq. (\ref{diagn}) and transforming this equation to an integral,
we obtain

\begin{eqnarray}
\label{tilden}
&&\tilde n_i(\bar t)
=\kern-4pt \exp  \left( \frac{t-\bar t}{\tau_i}  \right)
\left\{\tilde n_i(t)
+\frac{1}{\tau_i}\int\limits_{t}^{\bar t} dt'
\tilde n^{eq}_i(\mbox{\small\bf
R}(t')) \exp \left( \frac{t'-t}{\tau_i} \right) \right. \\
&+&\kern-4pt \left.\sum_{k}\int\limits_{t}^{\bar t}dt'
\int\limits_{t}^{t'}dt''\Omega_{ik}(t',t'')
\exp \left( \frac {t''-\bar t}{\tau_{ik}}\right)
[\tilde n_k(t'')-\tilde n_i(t'')]\right\},
% \frac {d\tn_i(\bt)}{d\bt} &=&
%\sum_k\int_t^{\bt}dt'\Omega_{ik}(\bt, t')
%\exp\left(\frac{t'-\bt}{\tau_{ik}/2}\right)\left[\tn_k(t')
%-\tn_i(t')\right]\nonumber\\
%&-&\frac 1{\tau_i}\left[\tn_i(\bt)-\tn_i^{eq}(\bR(\bt))\right],
\end{eqnarray}
where
$$\Omega_{ik}(t,t')=\frac {2}{\hbar ^2} {\rm Re}\left\{
V_{ik}(\mbox{\small\bf R}(t))
V_{ki}(\mbox{\small\bf R}(t')) \exp
\left[i \int\limits_{t'}^{t}dt''\tilde\omega_{ki}(\mbox{\small\bf R}(t''))
\right]\right\}.$$
%$$\Omega_{ik}(\bt,t')=\frac {2}{\hbar^2}Re\left\{V_{ik}(\relR(\bt))
%V_{ik}(\bR(t'))\exp\left[i\int_{t'}^{\bt}dt''
%\omega_{ik}(\bR(t''))\right]\right\}$$.

The formal solution of Eq. (\ref{tilden}) is 
\begin{equation}
\label{nsolut}
\tilde n_i(t)=\tilde n^{eq}_i({\bf R}(t)) \left[ 1-\exp
\left( \frac {-\Delta t}{\tau_i} \right)\right]+
n_i(t) \exp \left( \frac {-\Delta t}{\tau_i} \right),   
\end{equation}
where
\begin{equation}
\label{ndynam}
n_i(\bar t)=\tilde n_i(t)+
\sum_{k}\int\limits_{t}^{\bar t}dt'
\Omega_{ik}(t',t')\frac{\sin[\tilde\omega_{ki}(\mbox{\small\bf R}(t'))
(t'-t)]}{\tilde\omega_{ki}(\mbox{\small\bf R}(t'))}
[\tilde n_k(t')-\tilde n_i(t')].
\end{equation}
In fact, Eqs. (\ref{nsolut}) and (\ref{ndynam}) present an integral equation
for $\tilde n_i(t)$.

\section{Results and discussion}

In this section, we present the results of calculating  the radial
friction coefficient and the kinetic energy losses as a function 
of the inter-nucleus distance for trajectories corresponding to DIC.
The initial projectile energy, atomic masses, and charges of the
colliding nuclei are the initial information used in the  calculations. 
The single-particle potentials of the colliding nuclei are taken in
the Woods-Saxon form  with the parameters $r_0$=1.18 fm and $a=0.54$ fm.
The characteristic time parameter $\Dt$ is taken to be  equal 
to (0.8--1.0)$\cdot 10^{-22}$ s.

The equations of motion for $\relR (t)$ and the single-particle 
occupation numbers $\tilde n_i(t)$ have been solved numerically,
transforming differential equations into finite difference equations
with the time step $\Dt$ and the initial conditions $R(0)$=20 fm and
$\tilde n_i(0)$=1 or 0   for occupied and unoccupied states of the 
noninteracting nuclei, respectively. Matrix elements  
$\Lambda _{PP^{\prime }}^{(T)}, ~~(\Lambda _{TT^{\prime }}^{(P)})$, and 
$\larg_{PT}$ have been calculated using the procedure developed in 
\cite{26,27}. 

The relative motion along the trajectory depends on the nucleus-nucleus 
interaction potential, which is determined by a double folding of the 
effective nuclear and Coulomb interactions of the nucleons with the 
nuclear densities  of the interacting nuclei. Because of  nucleon exchange 
and particle-hole excitations, the nuclear densities  of the colliding 
nuclei evolve during the reaction and the nucleus-nucleus potential changes 
correspondingly. This effect is included in our calculations. As an example 
of these calculations we present  the radial friction coefficient 
for the  $^{64}$Zn(440 MeV) +$^{196}$Pt reaction as a function of 
$R$ for an approach phase of the reaction (solid curve) in Fig. 1. 
The results of the classical model of D.H.E. Gross and Kalinowski 
\cite{33} (dashed curves with stars) are shown for comparison,
along with the results of the microscopic model  developed in \cite{15} 
which are obtained with a constant temperature (dashed curve without stars). 
The curves correspond to different temperatures which increase from 0.5 MeV 
(bottom curve) to 2 MeV (upper curve). It can be  seen that the difference 
between our results and the results of \cite{33} increases with the 
increase in the overlap of the colliding nuclei. Comparing our results with 
the results of \cite{15}, we can see  that  our radial friction  
coefficient coincides with those found in \cite{15} for temperatures 
increasing as $R$ decreases. This kind of behavour of the radial friction 
coefficient obtained in \cite{17}. It is natural since it corresponds  
to an increase  in the kinetic energy loss during  the approach phase of 
the reaction. Unfortunately, the results for larger values of $R$ are not 
presented in \cite{15}. Thus, our calculations demonstrate the importance 
of  inclusion of the  time dependence of the single-particle occupation 
numbers in the calculations.

In Figs.2 and 3, the radial friction coefficients $\gamma_{RR}$ are given 
as a function of $R$ for both approach (solid curve) and departure  
(dashed curve) stages of the  $^{64}$Zn(440 MeV) +$^{196}$Pt and
$^{56}$Fe(480 MeV) +$^{208}$Pb reactions. It can be seen that 
$\gamma_{RR}$ takes a larger value for the approach stage compared
 to the departure  stage. For the second reaction, this difference 
 is significant.

In Fig. 4 and 5, the dependence of the total kinetic energy losses
as calculated by our method (solid curve) and as based on the classical 
model of \cite{33} realized with the code TRAJEC \cite{34}
is demonstrated for the  $^{64}$Zn(440 MeV) +$^{196}$Pt and
$^{56}$Fe(480 MeV) +$^{208}$Pb reactions, respectively. 
The solid and dashed arrows indicate the  moment corresponding to the
turning point of the trajectory in our calculations and in the 
classical model, respectively. It can be seen that, in contrast to the 
results of the classical model, where the kinetic energy is dissipated 
during a short time interval of the order of 0.4 $\cdot 10^{-21}$ s
at the very beginning of the  reaction,  this process takes  significantly 
larger time in our model. This result is a consequence of the smaller 
value of our friction coefficient relative to those used in the classical 
phenomenological models. As  seen in Fig.1, the friction coefficient 
obtained in \cite{33} significantly exceeds our friction coefficient 
where the interacting nuclei strongly overlap. Nevertheless, our friction 
coefficient reproduces the total amount of  kinetic energy loss 
observed  experimentally. If the results of the classical model 
calculations support the idea of the fast kinetic energy loss and 
thermalization, our calculations  support the idea of  gradual 
kinetic energy dissipation as was argued in \cite{3,35}. 

In Fig. 6, we show a correction to the nucleus-nucleus interaction
potential connected with the rearrangement of  nuclear densities 
during the approach phase of the reaction for the collision 
of $^{64}$Ni(320 MeV) + $^{208}$Pb. It can be seen that the correction 
increases in absolute value  with a decrease  of the internucleus 
distance $R$. For the same reaction, the nucleus-nucleus interaction 
potential found in the sudden approximation (dotted line) is shown 
in Fig. 7 with the dynamical correction (dashed line), which is 
discussed just above, and the kinetic energy as a function of the
trajectory (solid line).  We show that the correction to the sudden 
approximation produced by a rearrangement of the  particle 
distribution in the interacting nuclei is important and must be included 
in the  consideration.

 \section{Summary}

In conclusion, we have calculated  the  friction coefficient for DIC based 
on the microscopic model of the structure of the colliding nuclei and 
avoiding the assumption of a fast statistical equilibrium of the  dissipated 
kinetic energy. Our results demonstrate the  importance  of considering 
the friction coefficient as an exact dynamic function of the single-particle
occupation numbers without the  assumption of the fast statistical 
equilibrium of the excitation energy. Based on this result, we
have demonstrated that the kinetic energy dissipation in DIC is a
gradual process which takes up a significant part of a reaction time.

\acknowledgments 

We are grateful to Dr. H. Hofmann for the valuable discussions and
suggestions which stimulated the writing of this paper, and we also 
want to thank Dr. N.V. Antonenko for fruitful discussions. 
 We   thank  Dr. Zh.  Kurmanov 
and Ms. Ann Schaeffer for their assistance in preparing this
manuscript. The authors (G.G.A., R.V.J. and A.K.N.) are grateful to the
International Science Foundation  (Grant No. RFJ--000) and the Russian 
Foundation of Basic Research (Grant No. 95-02-05684) for financial 
support. \\

\appendix
\section{}

The value of $\tau _i$ is calculated using the results of the theory of 
quantum liquids \cite{30,31} 
\begin{eqnarray}
\label{deftau}
\frac {1}{\tau^{(\alpha)}_i}&=
&\frac {\sqrt{2}\pi}{32\hbar\varepsilon^{(\alpha)}_{F_K}}
\biggl[(f_K-g)^2+\frac{1}{2}(f_K+g)^2\biggr]
 \biggl[\Bigl(\pi \Theta_K\Bigr)^2 +
\Bigl(\tilde {\varepsilon}_i - \lambda^{(\alpha)}_K\Bigr)^2\biggr]
\nonumber\\
&\times&
\biggl[1+{\rm exp}\Bigl
 (\frac{\lambda^{(\alpha)}_K-\tilde {\varepsilon}_i}
{\Theta_K}\Bigr)\biggr]^{-1}, 
\end{eqnarray}
where 
$${\Theta}_K(t)=3.46\sqrt{\frac{E_K^{*}(t)}{<A_K(t)>}}$$
is the effective temperature determined by the amount of 
intrinsic excitation energy $E_K^{*}=E_K^{*(Z)}+E_K^{*(N)}$;
$<A_K(t)>=<Z_K(t)>+<N_K(t)>$, $\lambda _K^{(\alpha)}(t)$\,, 
and $E_K^{*(\alpha )}(t)$ are the mass number, chemical potential, 
and intrinsic excitation energies for the proton ($\alpha
=Z$) and neutron ($\alpha =N$) subsystems of the nucleus $K(K=P,T)$,
respectively, (for details, see \cite{26}). Furthermore, 
the finite size
of nuclei and the available difference between the numbers of neutrons and 
protons need to use the following expressions for the Fermi energies 
\cite{31}:
\begin{eqnarray}
\varepsilon^{(Z)}_{F_K}=\varepsilon_F \biggl[1-\frac{2}{3}\bigl(1+2f'\bigr)
\frac{<N_K>-<Z_K>}{<A_K>}\biggr],
\nonumber\\
\varepsilon^{(N)}_{F_K}=\varepsilon_F \biggl[1+\frac{2}{3}\bigl(1+2f'\bigr)
\frac{<N_K>-<Z_K>}{<A_K>}\biggr]\,,
\end{eqnarray}
where  $\epsilon _F$=37 MeV, 
\begin{eqnarray}
f_K=f_{in}-\frac{2}{<A_K>^{1/3}}(f_{in}-f_{ex}), 
\nonumber\\
f'_K=f'_{in}-\frac{2}{<A_K>^{1/3}}(f'_{in}-f'_{ex})
\end{eqnarray}
and $f_{in}$=0.09, $f_{in}^{\prime }$=0.42, 
$f_{ex}$=-2.59, $f_{ex}^{\prime }$=0.54, $g$=0.7 are the constants of 
the effective nucleon-nucleon interaction.

\begin{figure}
\caption{The radial friction coefficient calculated according to the  
present model for the  approach phase of  the  $^{64}$Zn(440 MeV) +
$^{196}$Pt reaction as a function of  $R$ (solid curve). 
  For comparison   the results of the classical model of 
  D.H.E. Gross and H. Kalinowski [32] are shown 
  (dashed curve with stars) 
  and also the results of the microscopic model 
developed in [15]  
which are obtained with a constant temperature 
(dashed curves without stars). The curves correspond to different 
temperatures which increase from 0.5 MeV (bottom curve) to 2 MeV 
(upper curve).}
\label{1_fig} 
\end{figure}

\begin{figure}
\caption{The radial friction coefficients $\gamma_{RR}$ as a function
of $R$ for both the approach (solid curve) and departure (dashed curve) 
stages of the  $^{64}$Zn(440 MeV) +$^{196}$Pt reaction.}
\label{2_fig} 
\end{figure}

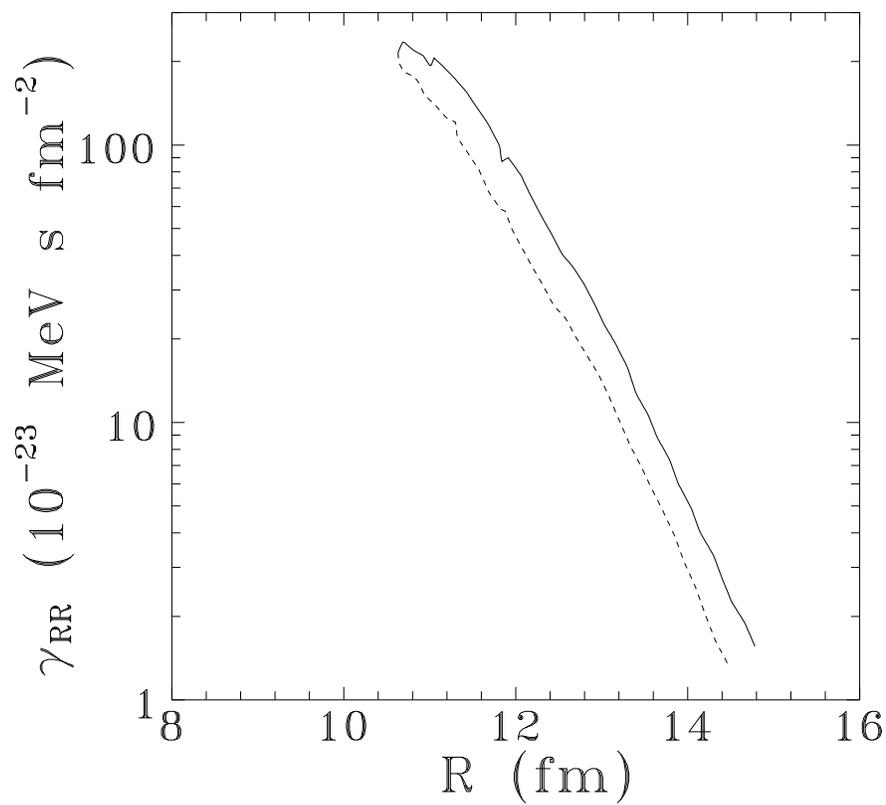
\begin{figure}
\caption{The same as in Fig.2 but for the 
  $^{56}$Fe(480 MeV) +$^{208}$Pb reaction.}
\label{3_fig} 
\end{figure}

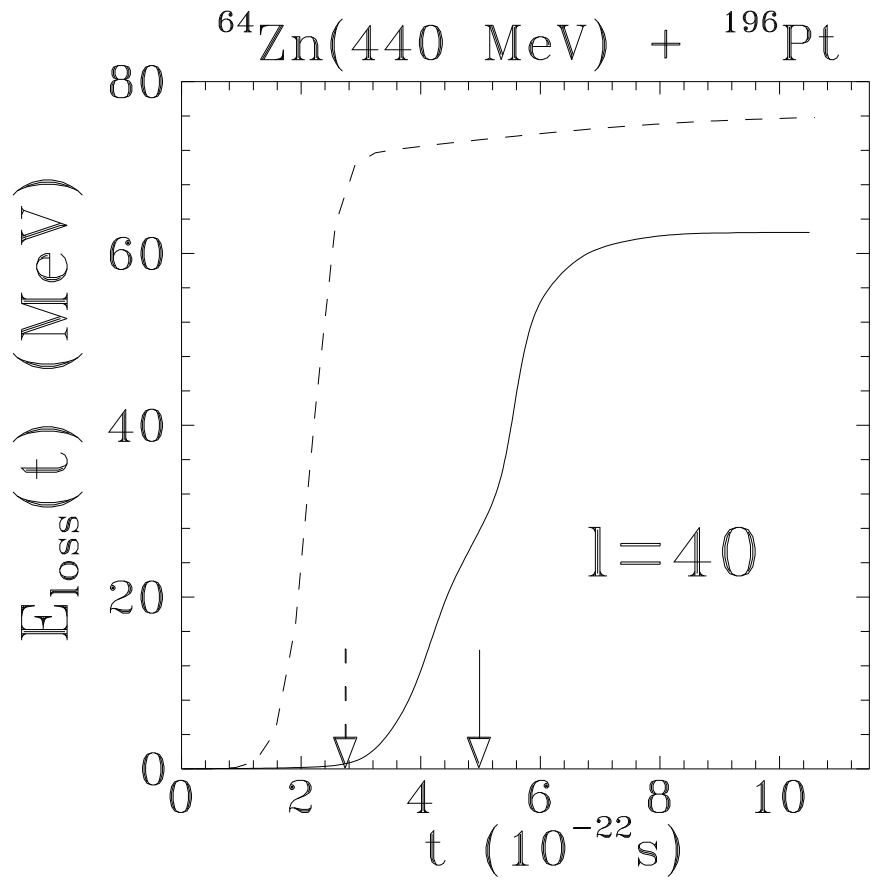
\begin{figure}
\caption{Dependence of the total kinetic energy losses
  calculated by our method (solid curve) and based on the classical 
model of [32]  
realized with  the code TRAJEC [33] for the  $^{64}$Zn(440 MeV) + 
$^{196}$Pt reaction. 
The solid and dashed arrows indicate the  moment corresponding to the 
turning point of the  trajectory in the present calculations and in the 
classical  model, respectively.}
\label{4_fig} 
\end{figure}

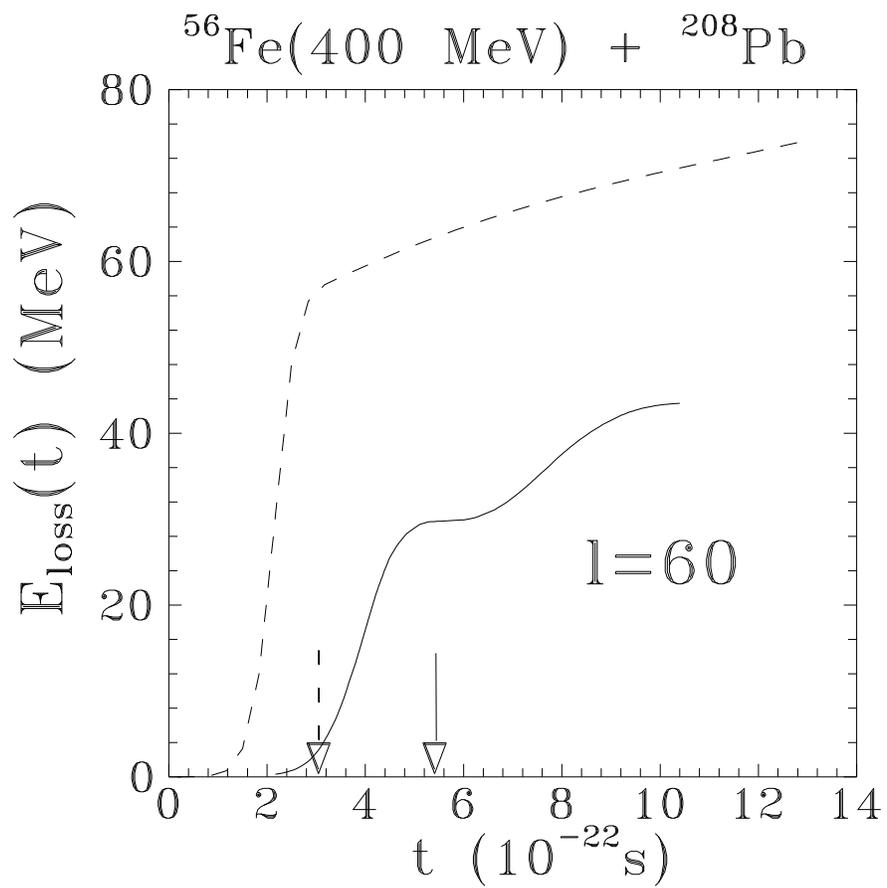
\begin{figure}
\caption{The same as in Fig.4 but for the 
  $^{56}$Fe(480 MeV) +$^{208}$Pb reaction.}
\label{5_fig} 
\end{figure}

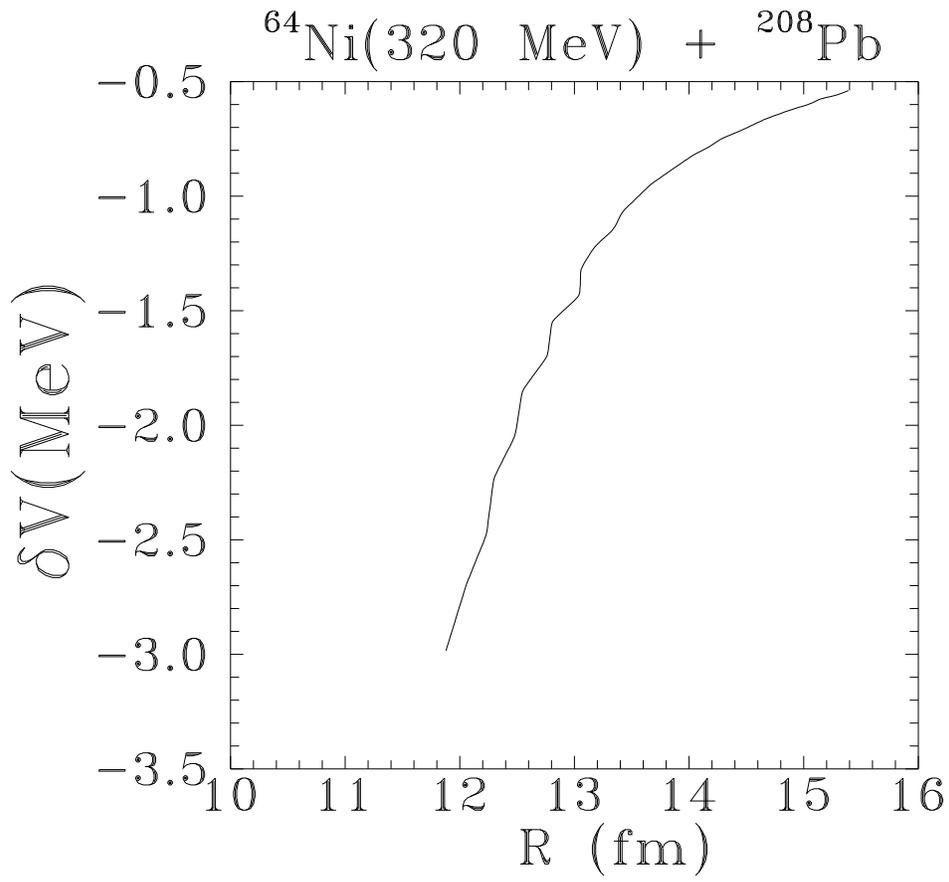
\begin{figure}
\caption{Correction to the nucleus-nucleus interaction
potential connected with a rearrangement of the nuclear densities during
the approach phase of the reaction for the collision 
of $^{64}$Ni(320 MeV) + $^{208}$Pb.}
\label{6_fig} 
\end{figure}

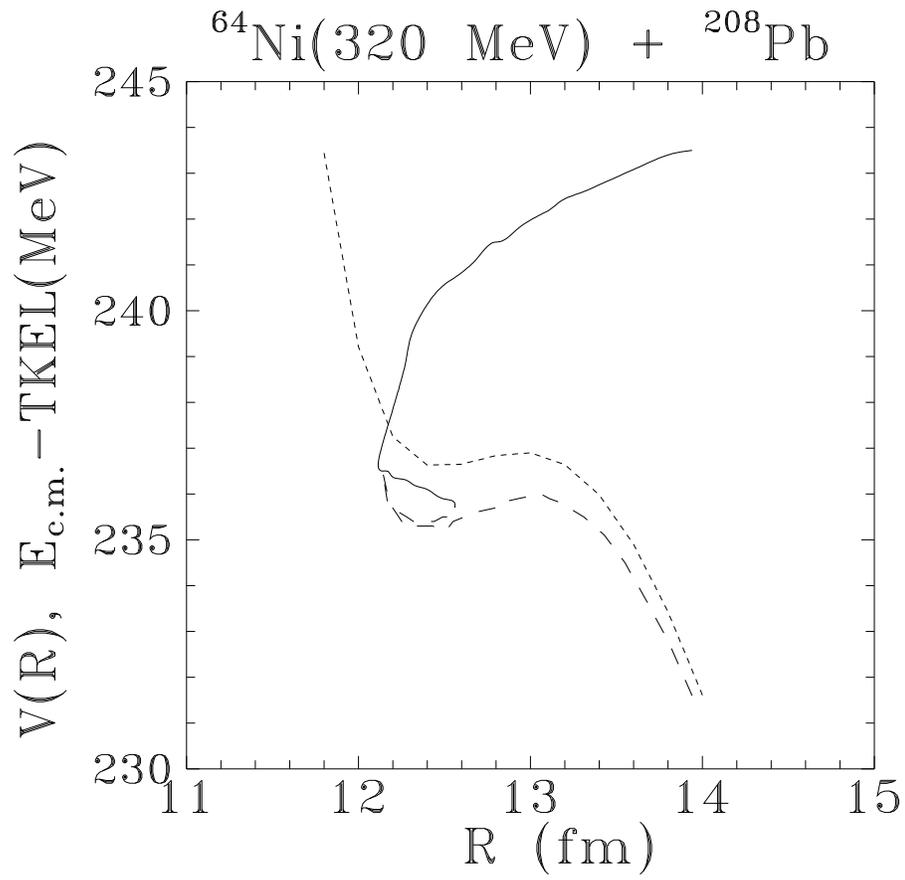
\begin{figure}
\caption{The nucleus-nucleus interaction potential found in the 
   sudden approximation (dotted line) and with the dynamic
correction  (dashed line) presented in Fig. 5, and 
the kinetic energy as a function of the  trajectory (solid line) for 
the  $^{64}$Ni(320 MeV) + $^{208}$Pb reaction.}
\label{7_fig} 
\end{figure}
\end{document}